\documentclass[smallextended]{svjour3}       
\smartqed  
\usepackage{graphicx,natbib
}
\usepackage{color}
\graphicspath{{Fig/}}
\begin{document}

\title{Size and shape-dependent melting mechanism of Pd nanoparticles
}

\author{Movaffaq Kateb \and Maryam Azadeh \and Pirooz Marashi \and Snorri Ingvarsson}
\institute{M. Kateb \at
              Science Institute, University of Iceland, Dunhaga 3, IS-107 Reykjavik, Iceland \\
              \email{mkk4@hi.is}          
           \and
           M. Azadeh \at
              Department of Mining and Metallurgical Engineering, Amirkabir University of Technology, Tehran, Iran
              \emph{Present address: School of Metallurgy and Materials Engineering, University of Tehran, Iran}
           \and
           	P. Marashi \at
              Department of Mining and Metallurgical Engineering, Amirkabir University of Technology, Tehran, Iran
           \and
           S. Ingvarsson \at
              Science Institute, University of Iceland, Dunhaga 3, IS-107 Reykjavik, Iceland
}

\date{Received: date / Accepted: date}

\maketitle

\begin{abstract}
Molecular dynamics simulation is employed to understand the thermodynamic behavior of cuboctahedron (cub) and icosahedron (ico) nanoparticles with $2-20$ number of full shells. The original embedded atom method (EAM) was compared to the more recent highly optimized version as inter-atomic potential. The thermal stability of clusters were probed using potential energy and specific heat capacity as well as structure analysis by radial distribution function ($G(r)$) and common neighbor analysis (CNA), simultaneously, to make a comprehensive picture of the solid state and melting transitions. The result shows ico is the only stable shape of small clusters (Pd$_{55}$ -- Pd$_{309}$ using original EAM and Pd$_{55}$ using optimized version) those are melting uniformly due to their small diameter. An exception is cub Pd$_{309}$ modeled via optimized EAM that transforms to ico at elevated temperatures. A similar cub to ico transition was predicted by original EAM for Pd$_{923}$ -- Pd$_{2075}$ clusters while for the larger clusters both cub and ico are stable up to the melting point. As detected by $G(r)$ and CNA, moderate and large cub clusters were showing surface melting by nucleation of the liquid phase at (100) planes and growth of liquid phase at the surface before inward growth. While diagonal (one corner to another) melting was dominating over ico clusters owing to their partitioned structure which retarded the growth of the liquid phase. The large ico cluster, using optimized EAM, presented a combination of surface and diagonal melting due to the simultaneous diagonal melting started from different corners. Finally, the melting temperature as well as latent heat of fusion were calculated and compared with available models and previous studies which showed, unlike the present result, the models failed to predict size-dependent motif crossover.
\end{abstract}

\section{Introduction}
Palladium nanoparticles have received growing interest due to their novel properties such as size \citep{chen2016} and shape (facet) \citep{wang2017} dependent catalytic activity, controlled oligomerization \citep{zhivonitko2016} and enhanced hydrogen storage \citep{rangel2016}. However, due to the higher surface to volume ratio \citep{schmitdt1998,safaei2008}, the nanoparticles are known as the most unstable structures among different nano-solids. Thus, it is necessary to predict their thermal stability for practical applications specifically when instability at elevated temperature can be considered as a failure. For instance, reshaping, melting or aggregation of catalyst clusters might happen far below the bulk melting point ($T_{mb}$) which results in changing their properties as well as functionality.

The nanoparticles melting temperature ($T_{mp}$) and enthalpy ($H_{mp}$), among the other thermodynamic properties, have received considerable attention \citep{qi2016}. Several models have been developed to predict the nanoparticles size-dependent $T_{mp}$ \citep{goldstein1992,jiang1999,safaei2010} and $H_{mp}$ \citep{zhang2000,jiang2002,attarian2008,fu2017}. However, developing a universal model requires an extensive effort and each model has to be verified with experimental or simulation data \citep{liang2017} which, to our knowledge, is not available for Pd clusters. In addition, these models fail to describe the melting dynamic and size-dependent melting mechanisms. For instance, quite recently, it has been shown that reshaping of Ag particles depends on the size and a transition from homogeneous to surface melting mechanism occurs by an increase in the particle size \citep{liang2017}.

Alternatively, atomistic Monte Carlo (MC) \citep{westergren2003} and molecular dynamics (MD) \citep{baletto2002,pan2005,miao2005,schebarchov2006} simulations have proven to be an excellent tool for probing the nanoparticles stability and melting behavior. For instance, \citet{baletto2002} have shown that for small Pd cluster icosahedron (ico) structure is more stable while decahedron (dec) and cuboctahedron (cub) are more stable at the moderate and large clusters, respectively. In this regard, they compared Rosato-Guillop{\'e}-Legrand and embedded atom method (EAM) force fields those were in close agreement. However, they only focused on the energetics of the different structures rather than showing a dynamic transition between allotropes. Thus, it still remains unclear that under what condition an unstable shape transforms into a more stable counterpart. Later, \citet{pan2005} reported in the case of Pd$_{309}$ cluster both of the cub and ico shapes are stable in a wide range of temperatures using Sutton-Chen (SC) potential. They also reported surface melting mechanism which was detected using radial distribution function, $G(r)$. However, the $G(r)$ has been defined separately for each shell of the cluster in their work which can be misleading i.e.\ the diffusion of the surface atoms, those have more degree of freedom, into inner shells causes a broadening of the $G(r)$ peak for the outer shell which can be misinterpreted as surface melting \citep{bertoldi2017}. They regretted this drawback by reporting quasi-solid core at the melting point ($T_{mp}$) of clusters while their $G(r)$ patterns indicated broad peaks for all shells \citep{pan2005}. They also observed a minor peak prior to the main melting peak in the specific heat capacity $C_p$ of the ico cluster which they attributed to the surface melting. Using a similar potential, \citet{miao2005} reported the surface melting for a spherical Pd$_{456}$ cluster based on bond-orientational order parameters (BOP). The BOP is a local structure characterization method obtained by determining the angle between a list of nearest neighbors (NN) \citep{steinhardt1983}. Thus, it determines which atoms are in solid/liquid states and enables the study of the surface melting more precisely. Later, \citet{schebarchov2006} pointed out the BOP is unable to distinguish between fcc and hcp \citep{steinhardt1983} and it is not efficient to detect solid state transitions in the clusters. To maintain brevity, cub to ico transition can be detected by mapping twining boundaries in ico cluster those can be characterized as hcp atoms. Thus, they suggested common neighbor analysis (CNA), instead of BOP, which is very sensitive to the angles between different pairs of NN \citep{tsuzuki2007}. {\color{blue}The efficiency of CNA for the structural transition has been demonstrated earlier by \citet{baletto2000,baletto2001} and interested reader is referred to Ref. \citep{rossi2007} to learn about the potential of CNA for the detailed characterization of clusters.} They successfully demonstrated dec to ico transition through a solid-liquid state for a Pd$_{877}$ cluster. This is interesting since they used the EAM force field and obtained a different result than static energy calculation by \citet{baletto2002} with similar potential. It is worth noting that for small clusters those have a high degree of freedom and might undergo several structure transformations at elevated temperatures, the Lindemann index \citep{lindemann1910} ($\delta_L$) gives scattered results as reported elsewhere \citep{alavi2006,zhang2013}.

In the present work, the phenomenological melting of symmetric shell Pd clusters of cub and ico structure is investigated. Since EAM potential predicts solid-state phase change, according to previous studies, original and highly optimized EAM were utilized in the present study. In addition, different methods of caloric curves, $G(r)$ and CNA were utilized to make a comprehensive understanding of the nanoparticles melting. This accompanied with a brief discussion on the limitation of each method and the resulting misinterpretation. Finally, the size-dependent results were compared to state of the art models.

\section{Method}
\subsection{MD simulation}
MD simulations were performed by solving Newton's equation of motion \citep{allen1989} using large-scale atomistic/molecular massively parallel simulator (LAMMPS) open source code\footnote[4]{version 14 Apr 2013 available at http://lammps.sandia.gov} \citep{plimpton1995,plimpton2012}. 

The original \citep{foiles1986} and highly optimized \citep{sheng2011} EAM was used to depict the inter-atomic potential between Pd atoms. Eq.~(\ref{eq:eam}) represents the original formulation of EAM potential:

\begin{equation}
    E_i=F_i\sum_{i\neq j}\rho_{ij}(r_{ij})+\frac{1}{2}\sum_{i\neq j}U_{ij}(r_{ij})
    \label{eq:eam}
\end{equation}
where $E_i$ and $F_i$ are cohesive and embedding energies of atom $i$, respectively. $\rho_{ij}(r_{ij})$ is the electron density of $j$ atoms located around the $i$ atom at distance $r_{ij}$. Clearly, $F_i$ is a many-body interaction term while $U_{ij}$ takes the pair interaction into account. 

The EAM potential was confirmed for describing solid characteristics such as cohesive energy and elastic constant as well as metals melting point \citep{foiles1986,sheng2011}. Moreover, it is reliable in determining the transitional properties especially the heat of fusion and heat capacities above room-temperature. The EAM has also been verified for a quantitatively correct description of such nanoscale systems; for instance, surface energy and geometry of low index surfaces.

The time integration of the equation of the motion was performed regarding the Verlet algorithm \citep{verlet1967,kateb2012} with a timestep of 3~fs. The temperature control was done using the Nose-Hoover thermostat with a damping of 30~fs. These conditions designed to generate positions and velocities sampled from canonical (NVT) ensemble. The initial velocities of the atoms were defined randomly from a Gaussian distribution at the appropriate temperature of 300~K and relaxed for 300~ps in NVT ensemble. Since practical nanoparticle melting is mostly performed in the vacuum, the heat associated with the particle's melting cannot be removed so efficiently by the surrounding medium. But the particles are in contact with the substrate which allows control over temperature and hence, the NVT ensemble provides a realistic representation of such systems. The simulations were performed by starting at 300~K, and then the temperature was elevated at a heating rate of $1.4\times10^{12}$~K/s.

\subsection{Cluster preparation}
\begin{figure}
    \centering
    \includegraphics[width=1\linewidth]{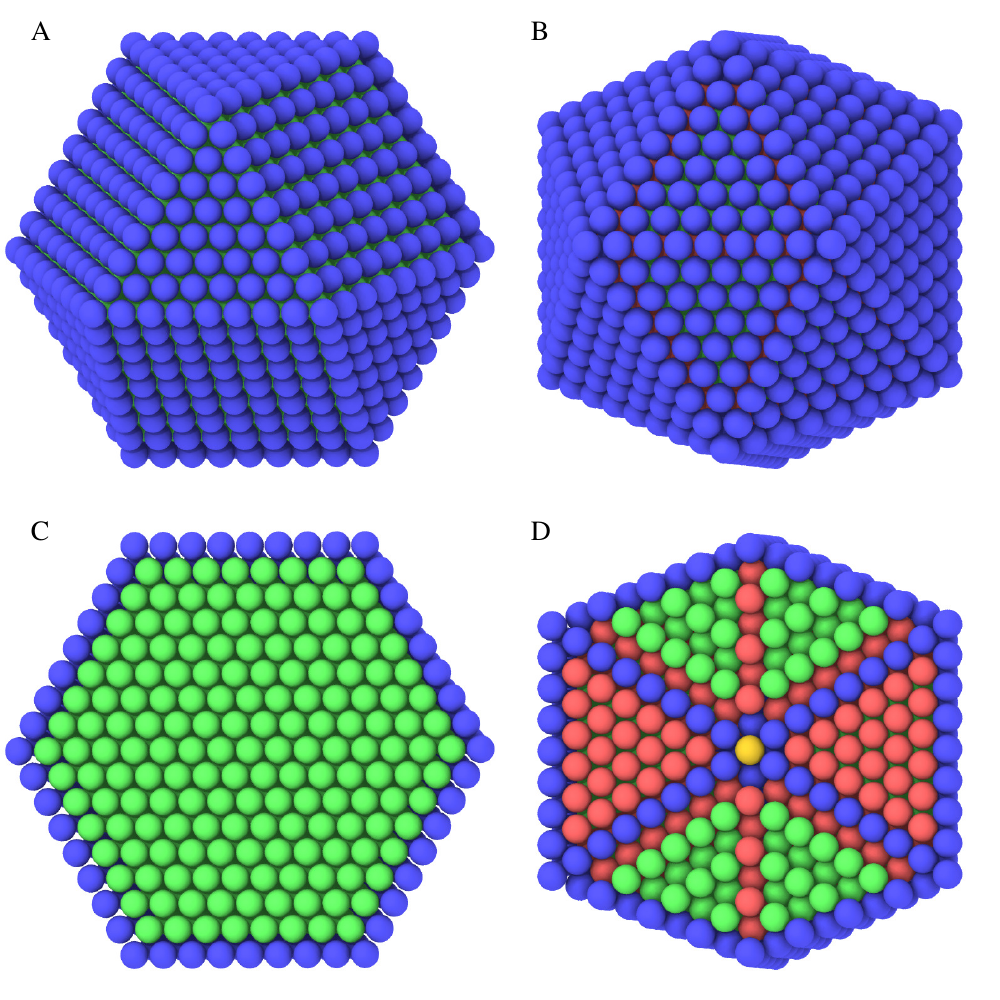}
    \caption{Illustration of (a) 8-cub and (b) 8-ico particles with 8 being the shell number, $n$. The crystal arrangement is shown through a slice of (c) 8-cub and (d) 8-ico indicating fcc, hcp, ico and unknown using green, red, yellow and blue colors, respectively.}
    \label{fig:cubico}
\end{figure}

The Pd nanoparticles were considered to be in the cub and ico forms those found in experimental characterizations \citep{jose2001}. For different sizes of given clusters, cub and ico were made considering the so-called magic number \citep{poole2003} or a total number of cluster atoms ($N_t$) which is described as the function of the shell number ($n$):

\begin{equation}
    N_t=\frac{1}{3}(10n^3+15n^2+11n+3)
\end{equation}
where $n=0$ denotes a mono-atomic system and $n\geq1$ defines the full shell clusters.
Here, clusters with sizes $\sim$1.5 -- 12~nm including $n=2-20$~shells ($N_t=55-28741$~atoms) were considered. Fig.~\ref{fig:cubico} illustrates the 8-cub ($n=8$) and 8-ico as an example. The figure also shows a slice of particles indicating 8-cub is made by fcc and surface atoms while the 8-ico presents more complicated arrangement i.e.\ 20 fcc tetrahedral (green) those surrounded an ico atom (indicated by yellow in Fig.~\ref{fig:cubico}d)  separated by twinning grain boundaries detected as hcp (red).

We chose to use the cluster radius ($R_p$) defined by Guinier formula \citep{miao2005} which is nearly constant during surface diffusion, solid-state transition and at the beginning of melting.

\begin{equation}
    R_p=R_g\sqrt{\frac{3}{5}}+r_a
\end{equation}
where $R_g$ is particle gyration radius and $r_a$ is atomic radius equal to 0.137~nm. The term particle size hereafter is referred to as $D=2R_p$ calculated at ambient temperature. 

\subsection{Melting criteria}

The melting is commonly detected from isotherm transition in the so-called caloric curve(s) e.g. potential energy ($U$) plotted versus temperature ($T$) \citep{shim2002,zhao2001}. \citet{qi2001} defined $T_{mp}$ as the temperature with the maximum apparent heat capacity. In MD simulations, specific heat capacity at constant pressure ($C_p$) can be calculated by derivative of the $U$ average ($U_{ave}$) during heating:

\begin{equation}
    C_p=\frac{dU_{ave}}{dT}+\frac{3}{2}\Re
\end{equation}
here $\Re$ stands for the universal gas constant.

It is worth noting that the precision of $T_{mp}$ value calculated from $C_p$ is dependent on the number available data points ($U$ vs.\ $T$) and a limited number of sampling, which is usually the case in MC simulation, results in a huge error.

Several methods are used in MD simulations to identify the melting process based on atomic specifics. The first criterion proposed by Lindemann \citep{lindemann1910} who stated the melting of crystals occurs when the average amplitude of atomic vibrations is higher than a threshold value. The global $\delta_L$ is a system average of atomic quantity which shows a linear increase with temperature increment in solid-state and a step change due to the melting. However, most of the vibrations of the surface atoms in the small clusters, which have more degree of freedom, were assumed as melting behavior by this model \citep{alavi2006,zhang2013}. This is a serious issue since it may lead to misinterpreting of the surface melting instead of a solid state transition.
The most straightforward structure analysis is offered by $G(r)$ or pair correlation function \citep{iida1988}. It describes how density varies as a function of distance in a system of particles from a reference particle.

\begin{equation}
    G(r)=\langle4\pi r^2\rho_adr\rangle_T
\end{equation}
where $\rho_a$ is the atom numbers density, $r$ is the distance from reference particle and $dr$ determines the bin size. The angle brackets i.e.\ $\langle\rangle_T$ denote the time average at constant $T$. 

Here $G(r)$ was extracted by 300 ps relaxation of clusters in NVT ensemble at desired $T$ i.e.\ 300 -- 2500~K with 100~K steps. Then the $G(r)$ averaged out over the whole time with 10$^{-4}$~nm bin size without periodic boundary condition (PBC).

This results in a pattern of several peaks corresponding to number and distance of NNs which applies to a wide range of materials. The melting transition causes a variation in the density and can be detected by shifting and broadening of peaks in the $G(r)$ pattern. However complex solid-state transition at nanoscale such as cub to ico with constant coordination number and even distance, is very hard to determine with $G(r)$.

Recently, CNA has shown promising tool due to providing the possibility of a distinction between allotropic transitions and melting process. Briefly, the CNA determining local crystal structure based on the decomposition of 1st NNs obtained from $G(r)$ into different angles \citep{faken1994,tsuzuki2007}. It is quite sensitive technique to the symmetry of different pairs of bonds. Thus a twining grain boundary as the main difference of ico and cub cluster can be determined based on a slight angle difference between 1st NNs while it holds entire properties of an fcc atom.

\subsection{Visualization}
The open visualization tool (OVITO) package\footnote[5]{Version 2.7.1 available at http://ovito.org/} was used to generate atomistic illustrations \citep{stukowski2009}.

\section{Result and discussion}
\subsection{Potential}

Table~\ref{tab:eam} summarizes important values reproduced by optimized EAM potential in comparison with original EAM and experimental or ab initio results. The fcc to hcp transformation energy, $\Delta E$, is optimized to be equal to 0.02~eV/atom in agreement with thermodynamically assessed value at room temperature \citep{dinsdale1991} and more precise than 0.026~eV/atom obtained by original EAM \citep{foiles1986}. This value is very important for realization of ico to cub transition and stability of ico structure. It is believed the vacancy formation energy, $E_v$, is the elementary mechanism of melting and thus it is one of the most important parameter to capture realized melting. The optimized force field results in vacancy formation energy of 1.4~eV similar to value estimated from experimental melting point \citep{kraftmakher1970} and better than 1.44~eV obtained from original potential \citep{foiles1986}. The values for surface energy, $\gamma_{sv}$, in the optimized EAM predicts higher values than original EAM regarding all plains. The experimental value of the $\gamma_{sv}$ is estimated from contact angle representing an average over all planes, but it is still higher than both EAM values. Finally, the optimized EAM results in $T_{mb}$ value closer to experimental value \citep{rao1964}.

\begin{table}
    \caption{The values obtained by optimized EAM \citep{sheng2011} force field in comparison with experimental or ab initio results and original EAM\citep{foiles1986}. The $a_0$ denotes lattice parameter.}
    \label{tab:eam}
    \centering
    \begin{tabular}{ccccc}
         \hline
         Parameter&&experiment&EAM&EAM\\
         (unit)&&or ab initio&original&optimized\\
         \hline
         $T_{mb}$~(K)&&1828 \citep{rao1964}&1680&1750\\
         $a_0 (nm)$ &&0.389  \citep{ashcroft2005}&0.389&0.389\\
         &(100)&2000 \citep{tyson1977}&1370&1645\\
         $\gamma_{sv}$~(mJ/m$^2$) & (110)&2000&1490&1747\\
         &(111)&2000&1220&1529\\
         $\Delta E$ (eV/atom)&{fcc-hcp}&0.02 \citep{dinsdale1991}&0.026&0.02\\    
         $E_v$ (eV/atom)&&1.4&1.44&1.4\\
    \end{tabular}
\end{table}

\begin{figure}
    \centering
    \includegraphics[width=1\linewidth]{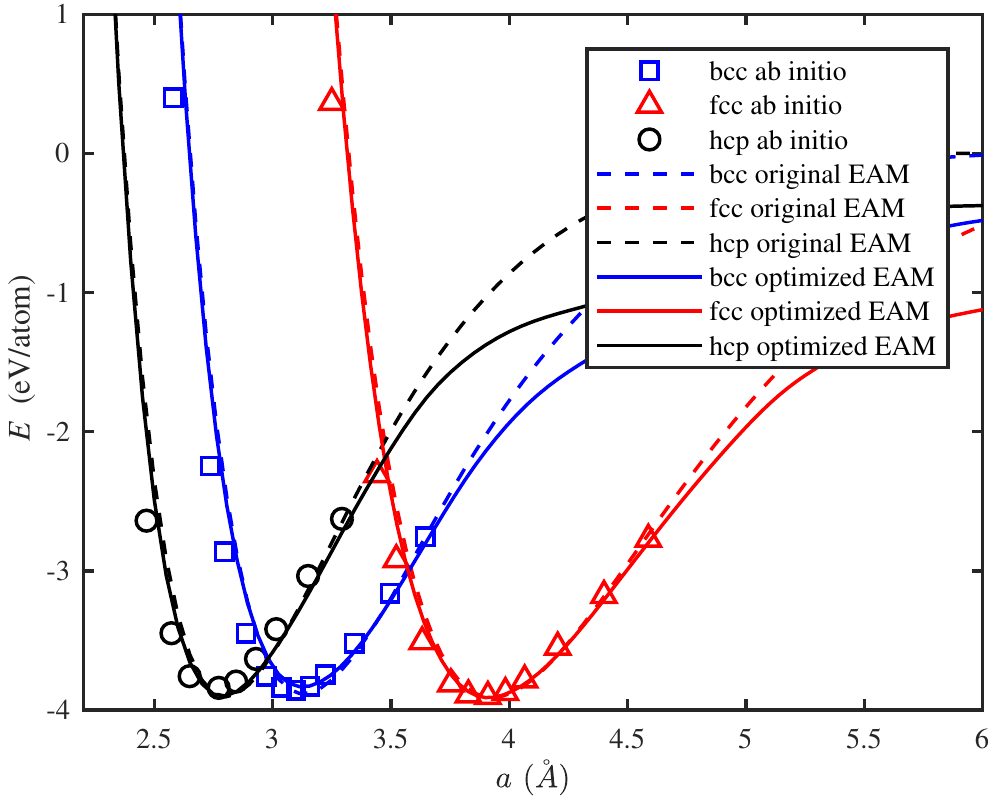}
    \caption{Validation of original and optimized EAM potentials for determining $E_c$ in comparison with ab initio calculation.}
    \label{fig:Ec}
\end{figure}

Fig.~\ref{fig:Ec} demonstrates the result of both EAM potential in determining the cohesive energy, $E_c$, of solid Pd in comparison with ab initio results. The minimum in fcc denotes equilibrium $a_0$ and $E_c$ equal to 0.389~nm and 3.911~eV/atom, respectively, which both EAM are in agreement with ab initio result. However, the $a_0$ and $E_c$ values for hcp determined by ab initio, are smaller than both EAM by 0.005~nm and 0.06~eV, respectively. In general, compared to ab initio results, both EAM are slightly off in the repulsive range, but they are acceptable in attraction part. The optimized EAM, however, tend to zero more smoothly.

Further, the dynamic structure factor, $S(q)$, was calculated which is a conventional method to examine liquid structure \citep{iida1988}. The $S(q)$ was calculated here using the Fourier transform of $G(r)$.
\begin{equation}
    S(q)=1+\rho_a\int(G(r)-1)e^{iqr}dr
\end{equation}
here $q$ is the wave-vector in reciprocal space.

The $G(r)$ was obtained in the same way described in subsection 2.3 but with  $dr=10^{-5}$~nm and PBC for 500 atoms in total. The wave-vector increment was chosen to be compatible with PBC i.e.\ $2\pi/L$ with $L$ being the cube side length equal to 5$\times a_0$ (1.945~nm).

\begin{figure}
    \centering
    \includegraphics[width=1\linewidth]{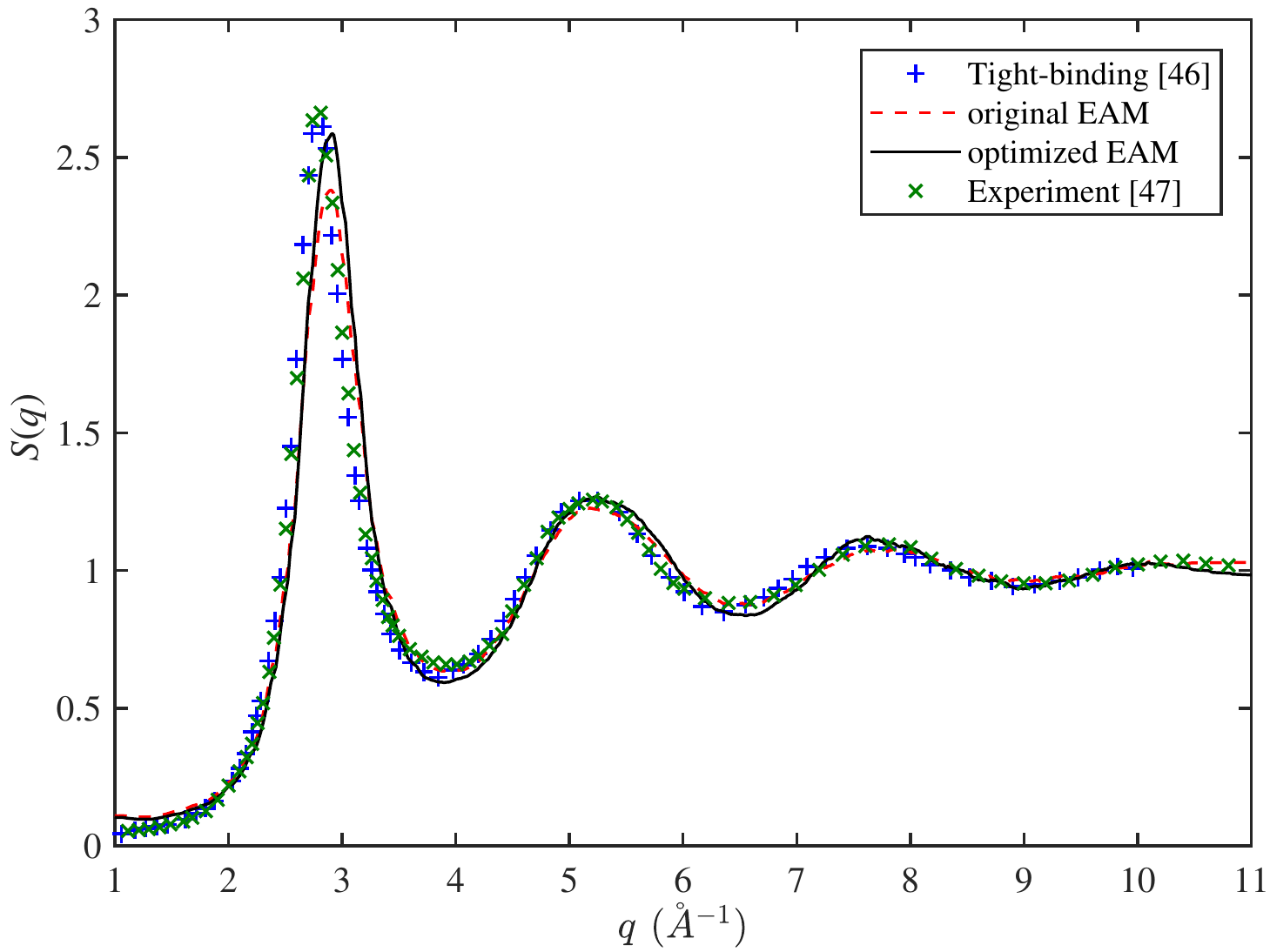}
    \caption{Validation of both EAM potentials for determining the $S(q)$ of molten Pd at 1853~K compared with results of tight binding \citep{alemany1999} and experiment \citep{waseda1980} at same temperature.}
    \label{fig:Sq}
\end{figure}

Fig.~\ref{fig:Sq} shows $S(q)$ for the liquid Pd calculated using both EAM in comparison with tight binding calculation \citep{alemany1999} and the experimental result obtained by x-ray Raman scattering \citep{waseda1980}. The figure clearly shows both EAM potentials are in close agreement with the tight binding and experimental results. However, the main discrepancies are the peaks' position predicted by both EAM those are slightly off. In addition $S(q)$ of the main peak obtained by original EAM shows notably smaller value (12\%).

\subsection{Relaxation}

Each cluster was relaxed as discussed in section 2.1 to minimize the potential energy of the entire system at the beginning. For the small cub clusters, a transformation to ico is expected during relaxation at room temperature \citep{baletto2002} The optimized EAM only predicts 2-cub transformation to 2-ico while relaxation using original EAM shows 2-cub and 4-cub are transforming to ico clusters in agreement with the previous result using the same potential \citep{schebarchov2006}. This difference originates from higher energy barrier for the cub to ico transformation in optimized potential. As mentioned in the introduction the solid state transition is expected to occur via an intermediate quasi-liquid state \citep{schebarchov2006} which seems to require higher energy using optimized EAM. In the following, we stick to ground state notations of the cluster and still call them 2-cub and 4-cub. {\color{blue}We would like to remark that the cub structure is not the best fcc structure with the lowest energy since its (100) facets are too large compared to (111) facets.}

The relaxation of higher shell number shows that there is no change in the shape and symmetry of the particle, indicating both potentials can successfully model the stable shapes of Pd clusters.

\subsection{Melting criteria}
\subsubsection{Caloric curves.}
\begin{figure}
    \centering
    \includegraphics[width=1\linewidth]{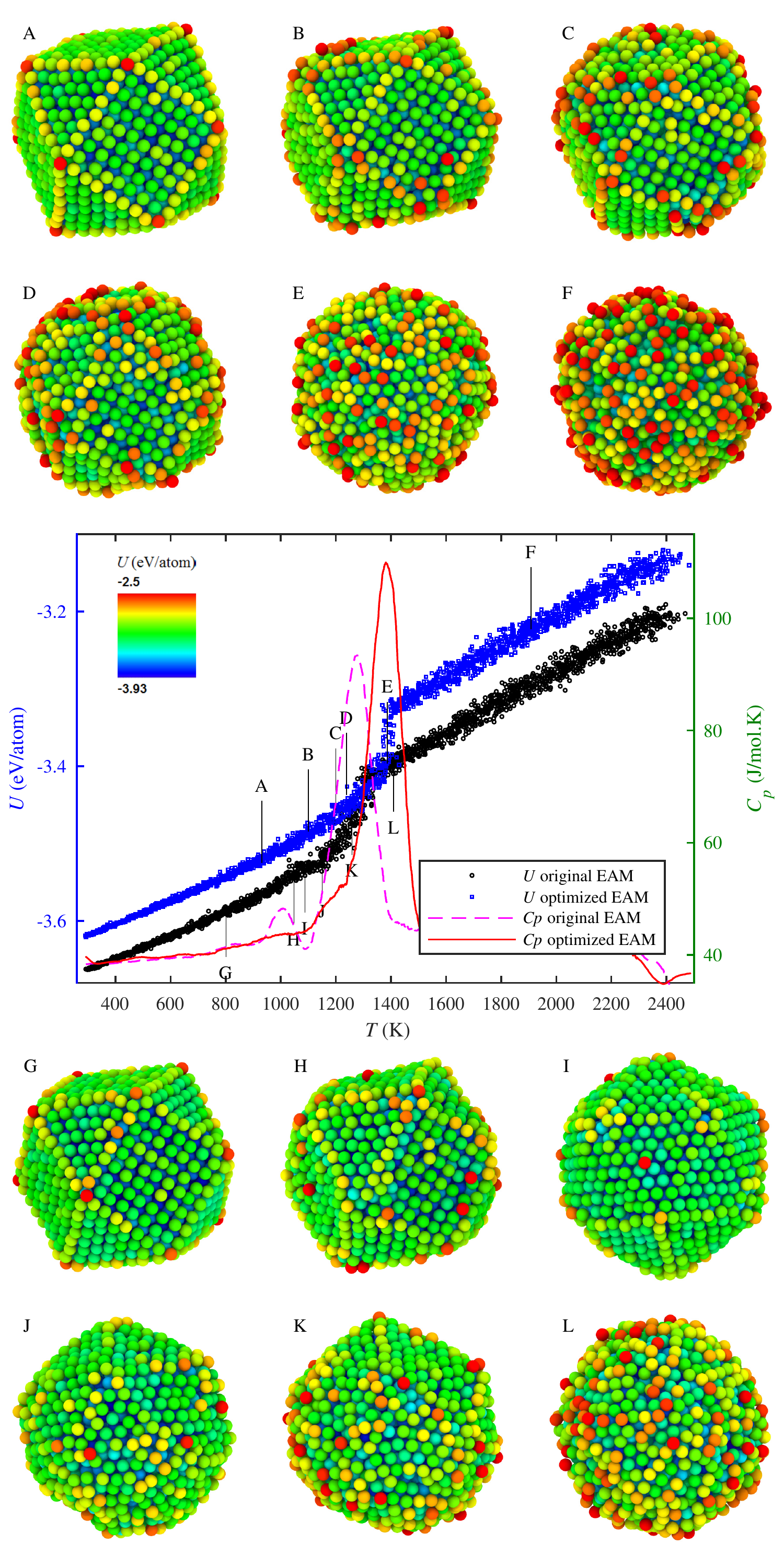}
    \caption{Variation of $U$ and $C_p$ with temperature for 8-cub particle using optimized and original EAM. (A -- F) are the corresponding snapshots of points A -- F obtained by optimized EAM while (G -- L) are snapshots of point G -- L from original EAM. The color bar in the inset indicates potential energy level of each atom.}
    \label{fig:caloric}
\end{figure}

Fig.~\ref{fig:caloric} illustrates the variation of $U$ and $C_p$ versus $T$ for the 8-cub particle (Pd$_{2057}$) and corresponding snapshots of point A -- F for optimized EAM and G -- L for original EAM. Both caloric curves present typical melting behavior including an isotherm transition due to $H_{mp}$. The gradual melting transition (non-isotherm melting) is more drastic in the curve obtained by original EAM which makes determining $T_{mp}$ harder. Nevertheless, $T_{mp}$ of 1277 and 1387~K were determined using main $C_p$ peaks for the original and optimized EAM, respectively. The difference came from the fact that optimized potential predicts bulk melting point more precisely and higher than original EAM. 
Snapshots A -- F belong to the optimized EAM which in general present more red atoms with higher potential energy. This is expected since $U$ for the optimized EAM is always stand above original EAM values. It can be clearly seen in snapshots A -- D, the cub structure remains unchanged without any evidence of surface melting but slightly rounding in the corners. Snapshots E and F were taken at and after $T_{mp}$ those are indicating a homogeneous melting mechanism. In snapshots G -- I a solid-state cub to ico transition is evident (see the the movie in the supplementary materials). This is associated with a clear step change in $U$ and corresponding local minimum in $C_p$ both at 1070~K from original EAM. It has been shown previously that, a solid-state transition is accompanied with a step change in $U$ and a local minimum in $C_p$, which was referred to as negative $C_p$ \citep{zhang2010}. The J -- L snapshots were taken close to melting point those showing a diagonal (corner to corner) melting which started at the bottom and moving up where (111) facet on the top can still be seen. This is the reason for drastic non-isotherm melting in $U$ of the original EAM. Finally, snapshot L shows a complete liquid state.

It is worth noting that the minor $C_p$ peak, is sometimes misinterpreted as surface melting \citep{pan2005}. However, as mentioned in the introduction, the improper use of $G(r)$ was the real reason for the misinterpretation. No surface melting is detected here as can be seen in snapshot H which is taken after minor $C_p$ peak.

\subsubsection{Radial  distribution function.}
\begin{figure}
    \centering
    \includegraphics[width=1\linewidth]{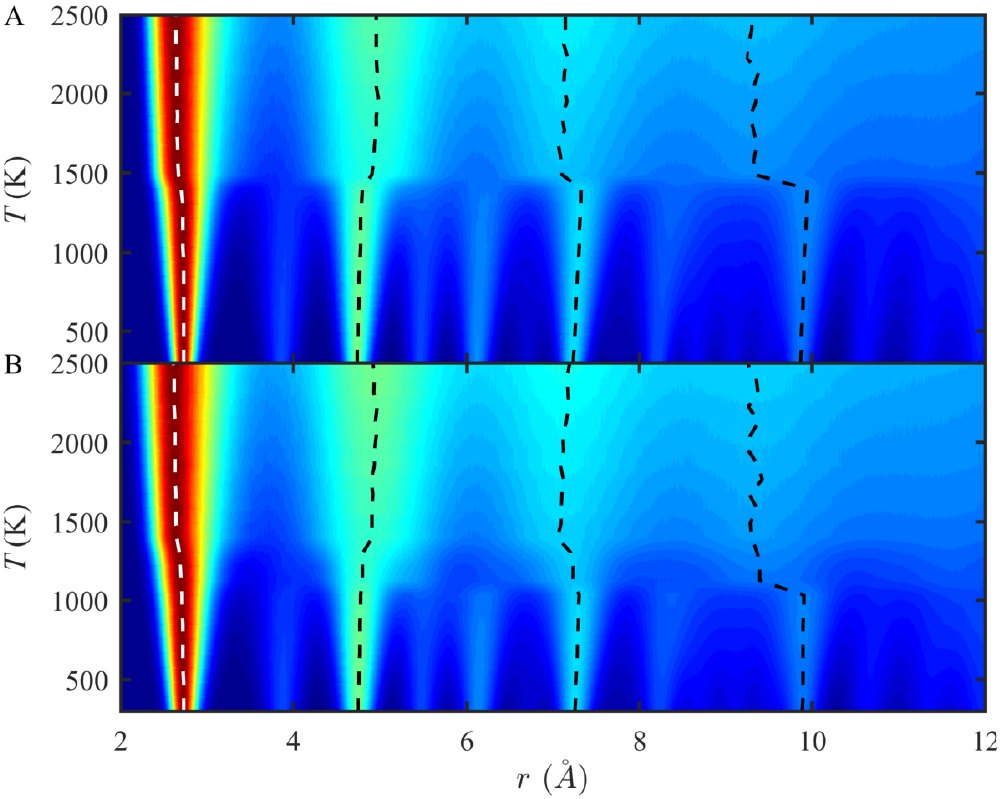}
    \caption{Variation of normalized $G(r)$ with $T$ for 8-cub cluster obtained by (a) optimized and (b) original EAM force fields. The colorbar indicates normalized $G(r)$ and dashed lines indicate its main peaks.}
    \label{fig:Gr}
\end{figure}

Fig.~\ref{fig:Gr} depicts the $G(r)$ variation with temperature for 8-cub using both potentials. In both cases, the $G(r)$ shows the same pattern for solid and liquid-like states in agreement with previously reported patterns \citep{pan2005}. The solid-like state at the bottom of each figure can be interpreted from 4 main peaks indicated by dashed line corresponding to 4 shells (not to be confused with 4th NNs). The 1st shell is consisting of 12 equidistant atoms and thus presents a single peak while further shells contain more peaks due to their geometrical complexity. The liquid-like state at top of each figure consisting of 4 broad peaks those are also indicated by dashed lines. It can be clearly seen that the increase in temperature causes peak broadening in both solid and liquid-like states. The melting occurs when there is step change in the peaks' positions (dashed lines). Fig.~\ref{fig:Gr}a shows homogeneous melting of all 4 shells at about 1400~K using the optimized EAM in agreement with 1387~K obtained from caloric curves. In Fig.~\ref{fig:Gr}b there are two step change in the peaks' positions at about 1000 and 1400~K. It is shown earlier that the first transition belongs to cub to ico transformation. However, since the step change in peaks' positions is more evident in the 4th peak (corresponding to 4th shell) it has been misinterpreted as shell by shell melting in the previous study \citep{pan2005}.

\subsubsection{Common neighbor analysis.}
\begin{figure}
    \centering
    \includegraphics[width=1\linewidth]{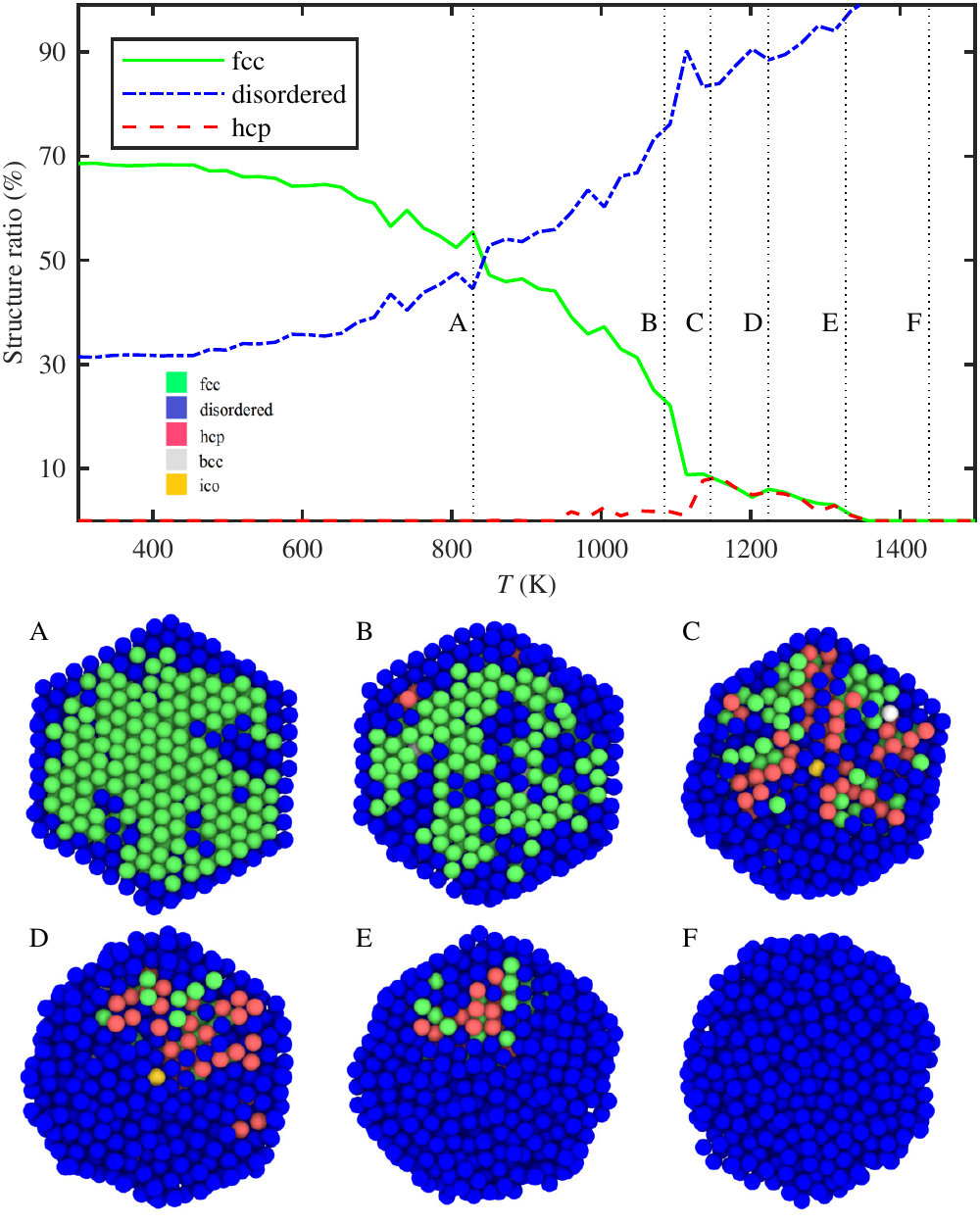}
    \caption{variation of structure ratios with $T$ and cross section of the corresponding snapshots at A -- F indicated in figure by dashed lines.}
    \label{fig:cna}
\end{figure}

Here the optimized EAM is neglected since original EAM present a more complicated case with a solid-state transition and drastic non-isotherm melting. Fig.~\ref{fig:cna} presents the variation in the ratio of different structures with $T$ according to the CNA for 8-cub cluster obtained by original EAM. The figure also includes corresponding snapshots of the particle cross section at points A -- F those indicated with dotted lines. The mirror change can be clearly seen in fcc and disordered structure, including surface atoms, till $\sim$940~K. The cross section of the particle at point A in this region shows only nucleation of disordered atoms close to the surface and remaining fcc atoms in green. In the 940 -- 1050~K range some hcp atoms are randomly appearing as can be seen in the snapshot B. It is worth noting that disordered phase exists everywhere even in the core part before transition to ico. Then the transition occurs at 1070~K, between point B and C, with a clear transition of fcc to disordered atoms ($\sim$13\%) followed by nucleation of hcp from disordered atoms ($\sim$7\%). Snapshot C shows an ico atom in the center (yellow) and the same portion of fcc and hcp (in red). It is worth noting that there are no fcc and hcp atoms in the small area at the bottom of the snapshot C which is associated with $\sim$6\% left disordered during the transition. Thus, CNA predicts an incomplete transition while we could not detect such thing using $G(r)$ and caloric curves. Thereafter both fcc and hcp transform to disordered atoms as shown in the D -- F snapshots. In specific, the diagonal melting is evident in D -- F with a disordered lower half growing towards crystalline aggregates at the upper half until total disorder is achieved.

The local maximum in the ratio of disordered atoms, between B -- C, was previously interpreted as ``coexistence of solid-liquid phase'' by focusing only on the CNA criteria \citep{schebarchov2006}. However, our $G(r)$ result shows no liquid-like state during this transition. Thus the transition occurs with a so-called military diffusion mechanism, which is very fast and short ranged, and the term ``transition quasi-liquid state'' is more preferred over solid-liquid coexistence. It is worth noting that the transition is limited to 6 and 8-cub clusters possibly because for the larger clusters, the activation energy required for this aim is higher than melting.

\subsection{Size and shape-dependent of the melting mechanism}

Table~\ref{tab:mech} summarizes melting mechanisms utilizing the above-mentioned criteria. Below 3~nm ($n=2-4$) the ico is the only stable shape of clusters before melting those are melting uniformly due to the small diameter of particles. For the larger diameters, cub clusters are generally melting with surface melting and ico clusters present diagonal melting. Optimized EAM predicts combined surface and diagonal melting for 16 -- 20-ico clusters ($10-12$~nm in diameter). It is worth noting that there is also a difference in the surface melting of cub and ico clusters. The surface melting of the cub clusters is associated with nucleation of the liquid phase at (100) planes followed by its growth at the surface which deforms polyhedral to a sphere with a lower surface area and surface energy (see the related movie in the supplementary materials). This is interesting since it is shown experimentally that the (100) planes present an incomplete surface melting in the bulk state while (110) and (111) planes present complete and no surface melting, respectively \citep{Vanselow1988}. Thus, the so-called two-stage melting is found to be associated with faster growth of liquid phase at the surface before moving towards the core. While in the case of large ico clusters, 16 -- 20-ico using optimized EAM, surface melting is in-fact simultaneous diagonal melting starting from different corners rather than formation of liquid shell followed by melting in the core (see the related movie in the supplementary materials). This difference can be explained by more resistance of (111) planes against surface melting compared to (100) ones as detected in many experiments. The interested reader is referred to reference \citep{Vanselow1988} and references therein. Thus, ico clusters with totally (111) planes at the surface present almost no surface melting. While cub cluster with both (100) and (111) planes at the surface shows surface melting. 

\begin{table}
    \caption{Dominant melting mechanism obtained using original \citep{foiles1986} and optimized \citep{sheng2011} EAM force fields. The superscript $^\star$ denotes cub to ico transition before melting and combined indicates simultaneous surface and diagonal melting mechanism.}
    \label{tab:mech}
    \centering
    \begin{tabular}{ccccc}
         \hline
         &\multicolumn {2}{c}{original EAM}&\multicolumn {2}{c}{optimized EAM}\\
         $n$&cub&ico&cub&ico\\
         \hline
         2&uniform&uniform&uniform&uniform\\
         4&uniform&uniform&uniform$^\star$&uniform\\
         6&diagonal$^\star$&diagonal&surface&diagonal\\
         8&diagonal$^\star$&diagonal&surface&diagonal\\
         10&surface&diagonal&surface&diagonal\\
         12&surface&diagonal&surface&diagonal\\
         14&surface&diagonal&surface&diagonal\\
         16&surface&diagonal&surface&combined\\
         18&surface&diagonal&surface&combined\\
         20&surface&diagonal&surface&combined\\
    \end{tabular}
\end{table}

\subsection{Size-dependent melting temperature}
\citet{safaei2010} developed a model considering the effects of the 1st NNs and the 2nd NNs atomic interactions. An approximation of the formula without considering of 2nd NNs atomic interaction is as follows:
\begin{equation}
    \frac{T_{mp}}{T_{mb}}=1-(1-q\frac{\bar{\epsilon}_s}{\epsilon_v})\frac{N_s}{N_t},q=\frac{\bar{Z}_s}{Z_v}
    \label{eq:Tm}
\end{equation}
here $N_s$ stands for the number of surface atoms, $q$ is the coordination number ratio with $Z_v$ equal to 12 for interior atoms and $\bar{Z}_s$ as the average coordination numbers of surface atoms. The $\epsilon_v$ and $\bar{\epsilon}_s$ respectively are bond energies of interior and surface atoms which the latter consists of cluster faces, edges and corners.

\begin{figure}
    \centering
    \includegraphics[width=1\linewidth]{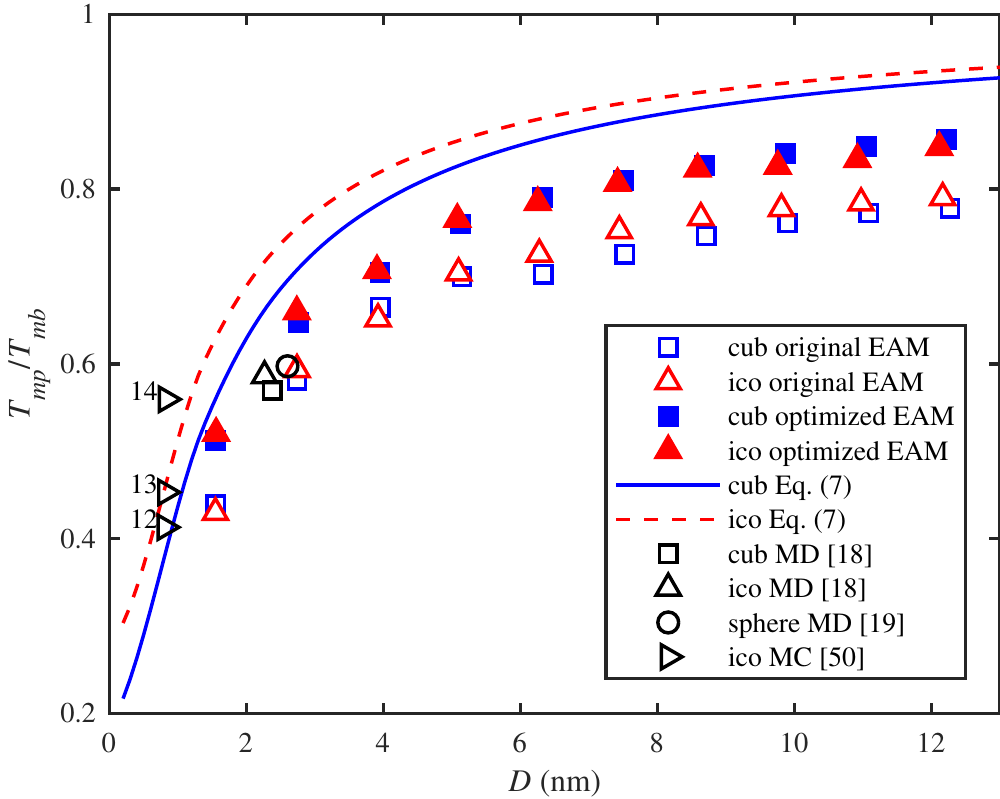}
    \caption{Dependence of the normalized $T_{mp}$ on the size of particle obtained from $U$ in comparison with MD \citep{pan2005,miao2005} and MC \citep{lee2001}  simulation as well as Safaei model \citep{safaei2010}. The entire data sets were normalized to $T_{mb}=1828$~K \citep{Vanselow1988}.}
    \label{fig:Tm}
\end{figure}

Fig.~\ref{fig:Tm} compares normalized $T_{mp}$ obtained from caloric curves in comparison with previous MD \citep{pan2005,miao2005} and MC \citep{lee2001} simulations. The figure also contains values calculated from Eq.~\ref{eq:Tm} assuming the bond strength for the surface and interior atoms to be equal. It is evident that $T_{mp}$ values obtained by optimized EAM always stands between original EAM and Safaei model. In addition, previous MD results using SC force field are predicting slightly higher values than original EAM but lower than optimized EAM. The original EAM data set shows a clear change around 5~nm for the cub due to the fact that 2 and 4-cub already transformed to ico during relaxation; and 4 -- 5~nm cub (corresponding to 6 and 8-cub) are transforming to ico before melting. In the case of optimized EAM, $T_{mp}$ is higher for ico till 5~nm and is lower for $D>$~5~nm. This is a clear indication of the stability of ico clusters till 5~nm and cub clusters for 6~nm and more. Thus the deficiency of the models is assuming a static lattice without showing crossover of different structure. This is because the model is focused on coordination number which is slightly lower for cub due to lower coordination number of (100) planes at the surface. While ico surface is only made out of (111) planes with higher coordination number.

The original EAM overestimates $T_{mp}$ for ico particles larger than 6~nm which means they are more stable than cub clusters. This might be associated with the $\gamma_{sv}$ difference in two potentials.

\subsection{Size-dependent melting enthalpy}
\citet{attarian2008} proposed the following model for the size-dependent $H_{mp}$ of clusters.
\begin{equation}
    \frac{H_{mp}}{H_{mb}}=[1-2(1-q)\frac{D_0}{D+D_0}].[1+\frac{3\Re T_{mb}}{2H_{mb}}\ln(1-2(1-q)\frac{D_0}{D+D_0})]
\end{equation}
with $D_0$ being specific diameter which the entire atoms are located at the surface (i.e.\ $N_s=N_t$) and $H_{mb}$ is the bulk melting enthalpy.

\begin{figure}
    \centering
    \includegraphics[width=1\linewidth]{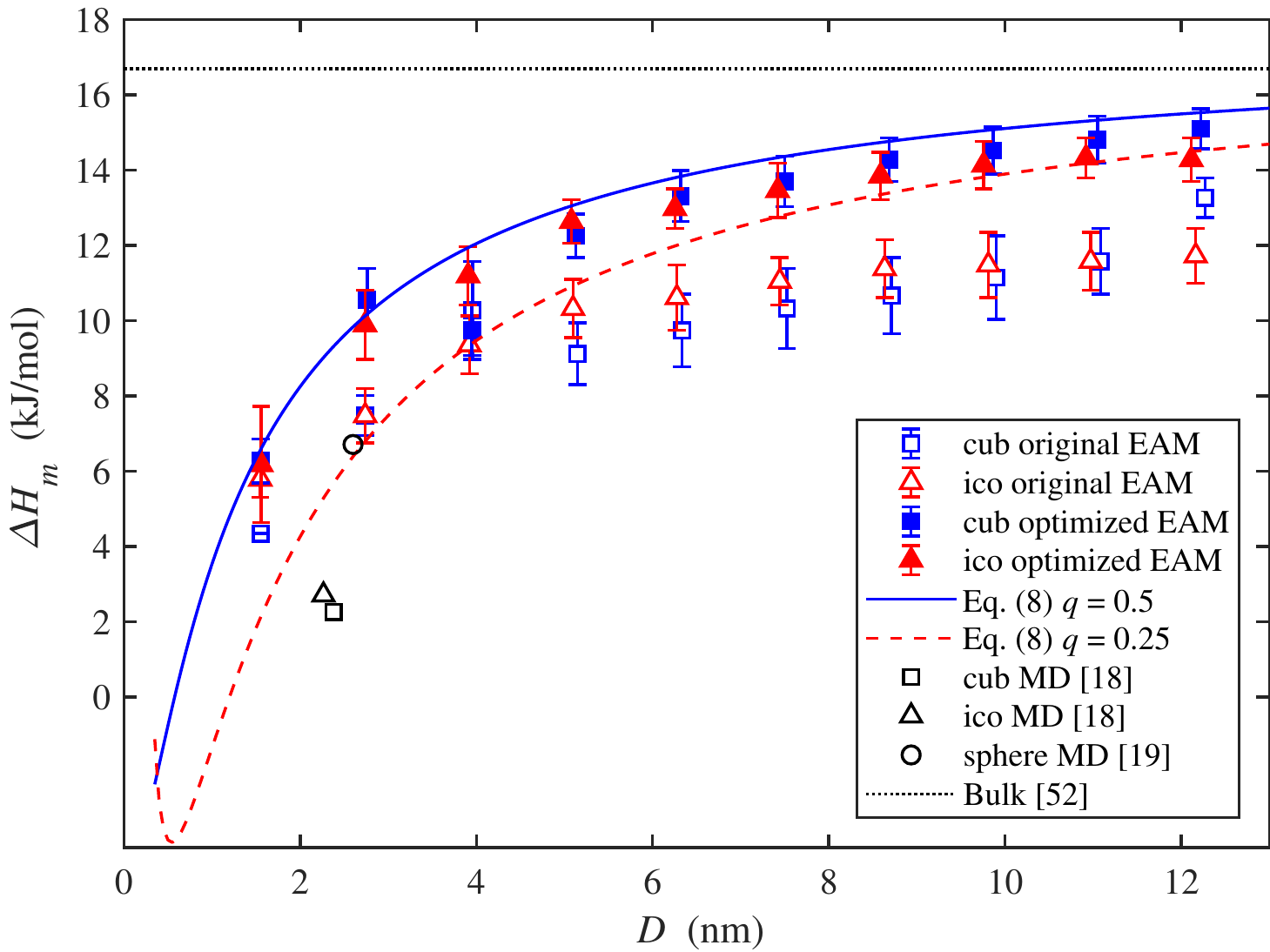}
    \caption{Variation of $H_{mp}$ with the diameter of Pd clusters obtained by optimized and original EAM potential, in comparison with previous MD results \citep{pan2005,miao2005} as well as \citet{attarian2008} model using $D_0= 0.6712$~nm and the bulk value 16.7~kJ/mol \citep{iida1988}.}
    \label{fig:dH}
\end{figure}

The variation of $H_{mp}$ with particle size is shown in Fig.~\ref{fig:dH} in comparison with \citet{attarian2008} model and previous MD results \citep{pan2005,miao2005}. It is worth noting that, the model shows a negligible difference between ico and cub and thus it plotted for different $\bar{Z}_s$ of 6 and 3 corresponding to $q=0.5$ and 0.25, respectively. As shown in the figure, the model predicts an increase in the $H_{mp}$ with the particle size in agreement with the result of the optimized EAM. While the original EAM underestimates $H_{mp}$ for particles larger than 5~nm. It can be seen SC potential \citep{pan2005} fails to predict correct values for cub and ico by a huge error ($\sim$100\%). While quantum corrected SC provides more accurate value for spherical cluster \citep{miao2005}. However, the model determines $H_{mp}<0$ below 0.5754 and 1.1984~nm respectively for $\bar{Z}_s$ of 6 and 3 meaning that melting smaller particles are exothermic and favorable. This failure originated from the crystalline basis of the model, which is not defined for a few atoms. Both EAM potentials predict a unified trend for ico particles. For the cub clusters, however, there is step change at $\sim$4~nm (corresponding to 6-cub) which is due to the transition to ico before melting. Thus the accuracy of models depends on knowing the stable structure of the cluster which they fail to predict.

\section{Conclusion}
In conclusion, the stability and melting behavior of Pd$_{55}$ -- Pd$_{28741}$ clusters of cub and ico shape studied by molecular dynamics simulation using two EAM force field i.e.\ original and highly optimized. Different melting criteria were discussed those make a clear picture of the melting process. The result shows small cub clusters (Pd$_{55}$ -- Pd$_{309}$) are unstable those are melting uniformly due to their small diameter. The only exception is Pd$_{309}$ using optimized EAM which transforms to ico at elevated temperatures before melting. A similar cub to ico transition is predicted by original EAM for Pd$_{923}$ -- Pd$_{2075}$ cluster and for larger clusters both cub and ico are stable up to the melting point. In general, as detected by $G(r)$ and CNA, the cub clusters present surface melting while ico clusters are melting diagonally (corner to corner) thanks to their partitioned structure. However, large ico cluster also present surface melting due to simultaneous diagonal melting from different corners. This is entirely different from surface melting observed in cub clusters. The later is associated with nucleation of liquid phase at the (100) planes and its growth at the surface before moving inward.

\subsection*{Conflict of Interest} The authors declare that they have no conflict of interest.

\bibliographystyle{spbasic}      
\bibliography{Ref.bib}   

\begin{thebibliography}{54}
\providecommand{\natexlab}[1]{#1}
\providecommand{\url}[1]{{#1}}
\providecommand{\urlprefix}{URL }
\expandafter\ifx\csname urlstyle\endcsname\relax
  \providecommand{\doi}[1]{DOI~\discretionary{}{}{}#1}\else
  \providecommand{\doi}{DOI~\discretionary{}{}{}\begingroup
  \urlstyle{rm}\Url}\fi
\providecommand{\eprint}[2][]{\url{#2}}

\bibitem[{Alavi and Thompson(2006)}]{alavi2006}
Alavi S, Thompson DL (2006) Molecular dynamics simulations of the melting of
  aluminum nanoparticles. The Journal of Physical Chemistry A 110(4):1518--1523

\bibitem[{Alemany et~al(1999)Alemany, Di{\'e}guez, Rey, and
  Gallego}]{alemany1999}
Alemany MMG, Di{\'e}guez O, Rey C, Gallego LJ (1999) Molecular-dynamics study
  of the dynamic properties of fcc transition and simple metals in the liquid
  phase using the second-moment approximation to the tight-binding method.
  Physical Review B: Condensed Matter 60(13):9208--9211,
  \urlprefix\url{https://link.aps.org/doi/10.1103/PhysRevB.60.9208}

\bibitem[{Allen and Tildesley(1989)}]{allen1989}
Allen MP, Tildesley DJ (1989) Computer simulation of liquids. Oxford university
  press, Oxford

\bibitem[{Ashcroft and Mermin(1976)}]{ashcroft2005}
Ashcroft NW, Mermin ND (1976) Solid State Physics. Holt, Rinehart and Winston,
  New York

\bibitem[{Attarian~Shandiz and Safaei(2008)}]{attarian2008}
Attarian~Shandiz M, Safaei A (2008) Melting entropy and enthalpy of metallic
  nanoparticles. Materials Letters 62(24):3954--3956

\bibitem[{Baletto et~al(2000)Baletto, Mottet, and Ferrando}]{baletto2000}
Baletto F, Mottet C, Ferrando R (2000) Reentrant morphology transition in the
  growth of free silver nanoclusters. Physical review letters 84(24):5544

\bibitem[{Baletto et~al(2001)Baletto, Mottet, and Ferrando}]{baletto2001}
Baletto F, Mottet C, Ferrando R (2001) Microscopic mechanisms of the growth of
  metastable silver icosahedra. Physical Review B 63(15):155,408

\bibitem[{Baletto et~al(2002)Baletto, Ferrando, Fortunelli, Montalenti, and
  Mottet}]{baletto2002}
Baletto F, Ferrando R, Fortunelli A, Montalenti F, Mottet C (2002) Crossover
  among structural motifs in transition and noble-metal clusters. The Journal
  of chemical physics 116(9):3856--3863

\bibitem[{Bertoldi et~al(2017)Bertoldi, Mill{\'a}n, and
  Guillermet}]{bertoldi2017}
Bertoldi DS, Mill{\'a}n EN, Guillermet AF (2017) Thermodynamics of the melting
  process in au nano-clusters: Phenomenology, energy, entropy and
  quasi-chemical modeling. Journal of Physics and Chemistry of Solids
  111:286--293

\bibitem[{Chen et~al(2016)Chen, Zhang, and Xu}]{chen2016}
Chen T, Zhang Y, Xu W (2016) Size-dependent catalytic kinetics and dynamics of
  {Pd} nanocubes: a single-particle study. Physical Chemistry Chemical Physics
  18(32):22,494--22,502

\bibitem[{Dinsdale(1991)}]{dinsdale1991}
Dinsdale AT (1991) {SGTE} data for pure elements. Calphad 15(4):317--425

\bibitem[{Faken and J{\'o}nsson(1994)}]{faken1994}
Faken D, J{\'o}nsson H (1994) Systematic analysis of local atomic structure
  combined with 3d computer graphics. Computational Materials Science
  2(2):279--286

\bibitem[{Foiles et~al(1986)Foiles, Baskes, and Daw}]{foiles1986}
Foiles S, Baskes M, Daw M (1986) Embedded-atom-method functions for the fcc
  metals {Cu}, {Ag}, {Au}, {Ni}, {Pd}, {Pt}, and their alloys. Physical Review
  B 33(12):7983

\bibitem[{Fu et~al(2017)Fu, Zhu, Xue, and Cui}]{fu2017}
Fu Q, Zhu J, Xue Y, Cui Z (2017) Size- and shape-dependent melting enthalpy and
  entropy of nanoparticles. Journal of Materials Science 52(4):1911--1918,
  \doi{10.1007/s10853-016-0480-9},
  \urlprefix\url{https://doi.org/10.1007/s10853-016-0480-9}

\bibitem[{Goldstein et~al(1992)Goldstein, Echer, and
  Alivisatos}]{goldstein1992}
Goldstein AN, Echer CM, Alivisatos AP (1992) Melting in semiconductor
  nanocrystals. Science 256(5062):1425--1427

\bibitem[{Iida and Guthrie(1988)}]{iida1988}
Iida T, Guthrie RIL (1988) The physical properties of liquid metals. Clarendon
  Press, Oxford

\bibitem[{Jiang et~al(1999)Jiang, Shi, and Zhao}]{jiang1999}
Jiang Q, Shi HX, Zhao M (1999) Melting thermodynamics of organic nanocrystals.
  The journal of chemical physics 111(5):2176--2180

\bibitem[{Jiang et~al(2002)Jiang, Yang, and Li}]{jiang2002}
Jiang Q, Yang CC, Li JC (2002) Melting enthalpy depression of nanocrystals.
  Materials Letters 56(6):1019--1021

\bibitem[{Jos{\'e}-Yacam{\'a}n et~al(2001)Jos{\'e}-Yacam{\'a}n,
  Mar{\'i}n-Almazo, and Ascencio}]{jose2001}
Jos{\'e}-Yacam{\'a}n M, Mar{\'i}n-Almazo M, Ascencio JA (2001) High resolution
  {TEM} studies on palladium nanoparticles. Journal of Molecular Catalysis A:
  Chemical 173(1):61--74

\bibitem[{Kateb and Dehghani(2012)}]{kateb2012}
Kateb M, Dehghani K (2012) Comparison of fracture behavior of sharp with blunt
  crack tip in nanocrystalline materials by molecular dynamics simulation.
  International Journal of Modern Physics: Conference Series 5:410--417

\bibitem[{Kraftmakher and Strelkov(1970)}]{kraftmakher1970}
Kraftmakher YA, Strelkov P (1970) Equilibrium concentration of vacancies in
  metals. In: Seeger A, Schumacher D, Schilling W, Diehl J (eds) Vacancies and
  Interstitials in metals: International Conference Proceeding, North Holland,
  Amsterdam, p~59, held in J\"ulich, Germany, September 23-28, 1986

\bibitem[{Lee et~al(2001)Lee, Lee, Kim, and Nieminen}]{lee2001}
Lee YJ, Lee EK, Kim S, Nieminen RM (2001) Effect of potential energy
  distribution on the melting of clusters. Physical review letters 86(6):999

\bibitem[{Liang et~al(2017)Liang, Zhou, Wu, and Shi}]{liang2017}
Liang T, Zhou D, Wu Z, Shi P (2017) Size-dependent melting modes and behaviors
  of {Ag} nanoparticles: a molecular dynamics study. Nanotechnology
  28(48):485,704

\bibitem[{Lindemann(1910)}]{lindemann1910}
Lindemann FA (1910) The calculation of molecular vibration frequencies.
  Physikalische Zeitschrift 11:609--612

\bibitem[{Miao et~al(2005)Miao, Bhethanabotla, and Joseph}]{miao2005}
Miao L, Bhethanabotla VR, Joseph B (2005) Melting of {Pd} clusters and
  nanowires: a comparison study using molecular dynamics simulation. Physical
  Review B 72(13):134,109

\bibitem[{Pan et~al(2005)Pan, Huang, Liu, and Wang}]{pan2005}
Pan Y, Huang S, Liu Z, Wang W (2005) Molecular dynamics simulation of
  shell-symmetric {Pd} nanoclusters. Molecular Simulation 31(14-15):1057--1061

\bibitem[{Plimpton(1995)}]{plimpton1995}
Plimpton S (1995) Fast parallel algorithms for short-range molecular dynamics.
  Journal of computational physics 117(1):1--19

\bibitem[{Plimpton and Thompson(2012)}]{plimpton2012}
Plimpton SJ, Thompson AP (2012) Computational aspects of many-body potentials.
  MRS bulletin 37(5):513--521

\bibitem[{Poole~Jr and Owens(2003)}]{poole2003}
Poole~Jr CP, Owens FJ (2003) Introduction to nanotechnology. John Wiley \&
  Sons, New Jersey

\bibitem[{Qi(2016)}]{qi2016}
Qi W (2016) Nanoscopic thermodynamics. Accounts of chemical research
  49(9):1587--1595

\bibitem[{Qi et~al(2001)Qi, {\c C}a{\v g}in, Johnson, and Goddard~III}]{qi2001}
Qi Y, {\c C}a{\v g}in T, Johnson WL, Goddard~III WA (2001) Melting and
  crystallization in {Ni} nanoclusters: The mesoscale regime. The journal of
  chemical physics 115(1):385--394

\bibitem[{Rangel et~al(2016)Rangel, Sansores, Vallejo, Hern\'andez-Hern\'andez,
  and L\'opez-P\'erez}]{rangel2016}
Rangel E, Sansores E, Vallejo E, Hern\'andez-Hern\'andez A, L\'opez-P\'erez P
  (2016) Study of the interplay between {N}-graphene defects and small {Pd}
  clusters for enhanced hydrogen storage via a spill-over mechanism. Physical
  Chemistry Chemical Physics 18(48):33,158--33,170

\bibitem[{Rao and Rao(1964)}]{rao1964}
Rao CN, Rao KK (1964) Effect of temperature on the lattice parameters of some
  silver-palladium alloys. Canadian Journal of Physics 42(7):1336--1342

\bibitem[{Rossi and Ferrando(2007)}]{rossi2007}
Rossi G, Ferrando R (2007) Freezing of gold nanoclusters into poly-decahedral
  structures. Nanotechnology 18(22):225,706

\bibitem[{Safaei(2010)}]{safaei2010}
Safaei A (2010) The effect of the averaged structural and energetic features on
  the cohesive energy of nanocrystals. Journal of Nanoparticle Research
  12(3):759--776

\bibitem[{Safaei et~al(2008)Safaei, Shandiz, Sanjabi, and Barber}]{safaei2008}
Safaei A, Shandiz MA, Sanjabi S, Barber ZH (2008) Modeling the melting
  temperature of nanoparticles by an analytical approach. The Journal of
  Physical Chemistry C 112(1):99--105, \doi{10.1021/jp0744681},
  \urlprefix\url{http://dx.doi.org/10.1021/jp0744681}

\bibitem[{Schebarchov and Hendy(2006)}]{schebarchov2006}
Schebarchov D, Hendy S (2006) Solid-liquid phase coexistence and structural
  transitions in palladium clusters. Physical Review B 73(12):121,402

\bibitem[{Schmidt et~al(1998)Schmidt, Kusche, von Issendorff, and
  Haberland}]{schmitdt1998}
Schmidt M, Kusche R, von Issendorff B, Haberland H (1998) Irregular variations
  in the melting point of size-selected atomic clusters. Nature
  393(6682):238--240

\bibitem[{Sheng et~al(2011)Sheng, Kramer, Cadien, Fujita, and Chen}]{sheng2011}
Sheng HW, Kramer MJ, Cadien A, Fujita T, Chen MW (2011) Highly optimized
  embedded-atom-method potentials for fourteen fcc metals. Physical Review B
  83(13):134,118,
  \urlprefix\url{https://link.aps.org/doi/10.1103/PhysRevB.83.134118}

\bibitem[{Shim et~al(2002)Shim, Lee, and Cho}]{shim2002}
Shim JH, Lee BJ, Cho YW (2002) Thermal stability of unsupported gold
  nanoparticle: a molecular dynamics study. Surface science 512(3):262--268

\bibitem[{Steinhardt et~al(1983)Steinhardt, Nelson, and
  Ronchetti}]{steinhardt1983}
Steinhardt PJ, Nelson DR, Ronchetti M (1983) Bond-orientational order in
  liquids and glasses. Physical Review B 28(2):784

\bibitem[{Stukowski(2009)}]{stukowski2009}
Stukowski A (2009) Visualization and analysis of atomistic simulation data with
  {OVITO}–the open visualization tool. Modelling and Simulation in Materials
  Science and Engineering 18(1):015,012

\bibitem[{Tsuzuki et~al(2007)Tsuzuki, Branicio, and Rino}]{tsuzuki2007}
Tsuzuki H, Branicio PS, Rino JP (2007) Structural characterization of deformed
  crystals by analysis of common atomic neighborhood. Computer physics
  communications 177(6):518--523

\bibitem[{Tyson and Miller(1977)}]{tyson1977}
Tyson WR, Miller WA (1977) Surface free energies of solid metals: Estimation
  from liquid surface tension measurements. Surface Science 62(1):267--276

\bibitem[{Vanselow and Howe(1988)}]{Vanselow1988}
Vanselow R, Howe RF (1988) Chemistry and physics of solid surfaces {VII},
  vol~10. Springer-Verlag Berlin Heidelberg, Berlin

\bibitem[{Verlet(1967)}]{verlet1967}
Verlet L (1967) Computer "experiments" on classical fluids. {I}.
  {T}hermodynamical properties of {L}ennard-{J}ones molecules. Physical Review
  159(1):98

\bibitem[{Wang et~al(2017)Wang, Xu, Zhao, Qi, Wang, Wang, Li, and
  Deng}]{wang2017}
Wang W, Xu J, Zhao Y, Qi G, Wang Q, Wang C, Li J, Deng F (2017) Facet dependent
  pairwise addition of hydrogen over {Pd} nanocrystal catalysts revealed via
  {NMR} using para-hydrogen-induced polarization. Physical Chemistry Chemical
  Physics 19(14):9349--9353

\bibitem[{Waseda(1980)}]{waseda1980}
Waseda Y (1980) The structure of non-crystalline materials: liquids and
  amorphous solids. McGraw-Hill International Book Co., New York

\bibitem[{Westergren and Nordholm(2003)}]{westergren2003}
Westergren J, Nordholm S (2003) Melting of palladium clusters–density of
  states determination by {M}onte {C}arlo simulation. Chemical physics
  290(2):189--209

\bibitem[{Zhang and Douglas(2013)}]{zhang2013}
Zhang H, Douglas JF (2013) Glassy interfacial dynamics of {Ni} nanoparticles:
  part {I} colored noise, dynamic heterogeneity and collective atomic motion.
  Soft matter 9(4):1254--1265

\bibitem[{Zhang et~al(2000)Zhang, Efremov, Schiettekatte, Olson, Kwan, Lai,
  Wisleder, Greene, and Allen}]{zhang2000}
Zhang M, Efremov MY, Schiettekatte F, Olson EA, Kwan AT, Lai SL, Wisleder T,
  Greene JE, Allen LH (2000) Size-dependent melting point depression of
  nanostructures: nanocalorimetric measurements. Physical Review B
  62(15):10,548

\bibitem[{Zhang et~al(2010)Zhang, Wen, Zhu, and Sun}]{zhang2010}
Zhang Y, Wen YH, Zhu ZZ, Sun SG (2010) Structure and stability of fe
  nanocrystals: An atomistic study. The Journal of Physical Chemistry C
  114(44):18,841--18,846

\bibitem[{Zhao et~al(2001)Zhao, Wang, Cheng, and Ye}]{zhao2001}
Zhao SJ, Wang SQ, Cheng DY, Ye HQ (2001) Three distinctive melting mechanisms
  in isolated nanoparticles. The Journal of Physical Chemistry B
  105(51):12,857--12,860

\bibitem[{Zhivonitko et~al(2016)Zhivonitko, Skovpin, Crespo-Quesada,
  Kiwi-Minsker, and Koptyug}]{zhivonitko2016}
Zhivonitko VV, Skovpin IV, Crespo-Quesada M, Kiwi-Minsker L, Koptyug IV (2016)
  Acetylene oligomerization over {Pd} nanoparticles with controlled shape: A
  parahydrogen-induced polarization study. The Journal of Physical Chemistry C
  120(9):4945--4953, \doi{10.1021/acs.jpcc.5b12391},
  \urlprefix\url{http://dx.doi.org/10.1021/acs.jpcc.5b12391}

\end{thebibliography}

%

\end{document}